\documentclass[a4paper,10pt]{article}

\usepackage{graphicx,amsmath,amssymb}
\usepackage{bm}
\usepackage{color}
\usepackage{setspace,textcomp}

\begin{document}
 
\title{Can topology reshape segregation patterns?}

\author{Yerali Gandica, Floriana Gargiulo and Timoteo Carletti \\
\emph{NaXys, University of Namur, Namur, 5000, Belgium}}

\date{\today}

\maketitle
\singlespacing
\begin{abstract}
We consider a metapopulation version of the Schelling model of segregation over several complex networks and lattice. We show that the segregation process is topology independent and hence it is intrinsic to the 
individual tolerance. The role of the topology is to fix the places where the segregation patterns emerge. In addition we address the question of the time evolution of the segregation clusters, resulting from different dynamical regimes of a coarsening process, as a function of the tolerance parameter. We show that the underlying topology may alter the early stage of the coarsening process, once large values of the tolerance are used, while for lower ones a different mechanism is at work and it results to be topology independent.
\end{abstract}
\newpage \baselineskip1.0cm
\singlespacing
\section{Introduction}
People tends to move to reduce their uneasiness and increase their personal utility computed across many dimensions, thus 
formation of strongly separated groups or even ghettos is a self-organised emergent process of our societies. Such phenomenon can be harmful when leading to discrimination based on segregation \cite{CutlerGlaeser1997}. This issue has been recognised as one of the most important socio-political problems in the USA, and many Western-European countries 
are becoming increasingly aware of this problem \cite{FagioloEtAl2007}. Economical inequalities have been recognised as the main cause for this phenomenon
\cite{FreyFarley1996}. Not downplaying this fact, Shelling studied, in the 70's, the effect of individual {preferences concerning the composition of the individuals neighbourhood}. {Introducing a stylised model, he showed} that even weak preferences unbalance can lead to total segregation in societies \cite{Schelling1971}. In the modern language of complex systems, such phenomenon can be ascribed to self-organisation in social sciences [8, 9, 10].

In the original works by Shelling, a population composed by two kinds of agents, say Red and Blue, is assumed to live on a
regular network (1D or 2D lattices) where each node can {either host} a single agent  or remain empty 
\cite{Schelling1969,Schelling1971}. Agents are happy, and thus do not move, if they are surrounded by sufficiently many 
agents of the same kind, otherwise they are unhappy and move to another location. {The model is specified though two 
parameters: the population density of agents on the lattice, $\rho$, and  the tolerance, $\varepsilon$, the parameter 
responsible for the willingness to move or to stay in a certain site.} Besides its simplicity the model exhibits a very rich phenomenology, for instance scholars have been interested in understanding the presence of a phase transitions in the final distribution of agents (segregation vs well mixed), the role of the density of agents and thus of the density of empty spaces (emptiness), and the dynamics of the separation of clusters composed by the same kind of agents (interface growth) \cite{VinkovicKirman2006,DallAstaEtAl2008}. 

Recently the Shelling model has been improved to account for the possibility that nodes {host} more than one agent \cite{RogersMcKane2012,DurettYuan2014,gargiulo}, opening in this way to a generalisation to the metapopulation framework \cite{Hanski1999}. In particular, in a previous work \cite{gargiulo}, we studied a metapopulation Schelling model over regular networks (1D and 2D lattices) able to capture the idealised movement of people across city districts, each one allowing for a finite carrying capacity. {A first result of this paper is that a phase transition happens when the tolerance is $\varepsilon_c=0.5$. For $\varepsilon<\varepsilon_c$ the lattice sites are completely magnetised, namely only one type of agents is present on each node.}
Moreover, we found that for low values of the tolerance ($\varepsilon<0.3$) the system remains stuck in a long quasi-stationary transient phase, during which the population is in a well mixed phase, but suddenly the system jumps in a new stable phase where 
agents are heterogeneously distributed across nodes. We named towers, such nodes with a very large density, corresponding 
thus to nodes where agents prefer to live. Let us observe that such result is obtained without any exogenous preferential
attachment mechanism. Another interesting characteristics exhibited by the
system in the range of small tolerances, is the 
presence of clusters of empty nodes (or links) separating mono-coloured
clusters; moreover such 
interfaces grow in time with a power law behaviour with a specific exponent
depending on the tolerance parameter. 

In the own Schelling words, people get separated along many lines and many ways, but the pattern formation regarding this separation has strong connection
with how different cultures define their satisfaction, in terms of their daily activity and living constraints. For instance, for families whose children 
play outside or love having their daily commercial life surrounding their houses, their neighbourhood is very important. On the other hand, if we 
consider cities, where people activity is largely restricted to the buildings where they live or work, then the individuals
relocation decisions depend more on the composition of the fully-connected building than on the neighbourhood. 

Building on the previous remark we are interested in the present work to consider a metapopulation Schelling model where
the underlaying network exhibits a complex topology and thus to understand how such heterogeneity of connectivity changes 
the model outcomes with respect to the regular lattice previously studied.

We find that a complex network structure does not change drastically the qualitative system behaviour and 
that the transition threshold leading to segregated states is $\varepsilon=0.5$ and thus results to be an universal 
parameter of the Schelling models. This outcome is interesting since it mirrors a previous important result for opinion 
dynamics, \cite{UniversalityBC}, where it has been showed that in Bounded Confidence Models, where a threshold determines
if people locally agree or not, the topology is not affecting the final outcome of the system, unless the network itself 
has a dynamics \cite{OpDynNetworks,OpDynHolme}. 

Even if the qualitative behaviour of the Schelling model is not affected by the underlaying network topology, we 
observe that the convergence times to the magnetised state depends strongly on the connectivity of the network, the larger
the average degree the longer the time needed to reach the final magnetised state. Moreover we observe important 
correlations between the local network structure and the population distributions, namely agents tend to avoid hubs, 
that thus will be mainly empty, while they tend to accumulate, namely to form towers, in low degree nodes. 

One the other hand, the removal and rewiring of local connections on nodes modifies the typical early coarsening 
process of clusters growth, for big values of $\epsilon$. However for low values of tolerance, when the system passes
for a long quasi-stationary regime, a late stage of coarsening becomes dominant, where the system creates the \lq\lq towers\rq\rq. We found that in this regime the time scaling is independent of the network topology. Our work is organised as follow: in section 2 we provide a detailed description of the methodology in terms of dynamics and topology. The results are shown in section 3. Finally we conclude in section 4.  

\section{The model} \label{sec:model}

Let us consider a network composed by $N$ nodes each one able to contain at most $L$ agents, some of which belong to 
the Red group and some to the Blue one. The total number of agents is a conserved parameter that we decide to express in terms
of the total emptiness, $\rho NL$, where $\rho\in(0,1)$, is defined as the total number of available space not occupied by
any agent. The system is initialised by choosing uniformly random amounts of Blue and Red agents in each node close to the homogeneous equilibria $(1-\rho) NL/2$. Time increases by discrete steps, one step consists 
in the random selection and the (eventual) relocation of an agent accordingly to her happiness. An agent is happy in a 
given node $i$, if the fraction of agents with the opposite kind inside the neighbourhood, defined as the set of nodes at 
distance smaller or equal to $1$ from $i$, is smaller than a threshold $\epsilon$, in formula:
\begin{equation}
\label{eq:happ}
f_i^B=\frac{\sum_{j\in i}n_j^A}{\sum_{j\in i}(n_j^A+n_j^B)}\text{ then $B$ is unhappy in node $i$, if $f_i^B>\epsilon$}\, ,
\end{equation}
where $n_i^X$, $X=A,B$, denotes the number of agents belonging to the $X$--kind in the node $i$--th, and $j\in i$ is a shorthand to 
denote the set of nodes connected to $i$ and $i$ itself. Unhappy agents relocate themselves to another node in the 
network, chosen with uniformly random probability, and provided that there is enough space available there.
  
In the following we consider networks formed of $N=400$ nodes and we will consider four topologies: Lattices, Small Word 
networks constructed using the Watts-Strogatz (WS) algorithm \cite{Watts_Stogatz}, Random Networks by using 
Erd\H{o}s-R\'enyi procedure \cite{ER} and the Scale-free topologies obtained using the preferential attachment mechanism
proposed by Barabasi-Albert (BA) \cite{BA}. To check the impact of the node degree we also build networks using the 
degree-preserving rewiring procedure proposed in \cite{maslov}.
 
\section{Results}

\begin{figure}[h]
\begin{center}
\includegraphics[width=11cm]{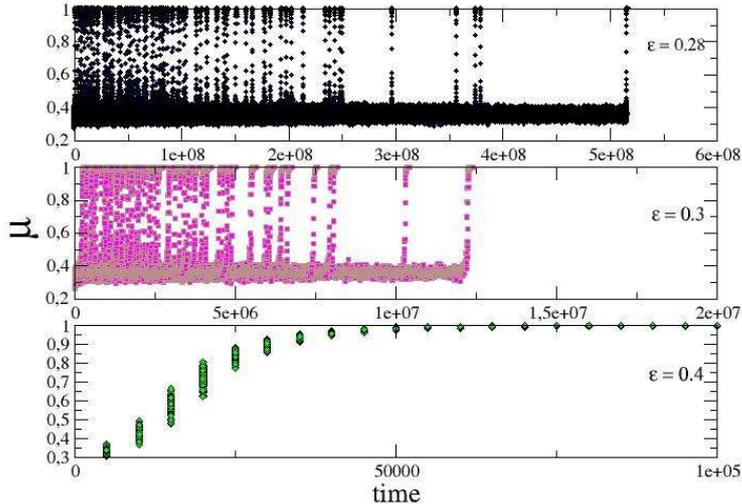}
\end{center}
\caption{Magnetization as a function of time for $100$ generic simulations performed using a scale free (BA) network
with $N=400$ nodes. We show results for $\epsilon=0.28$ (upper), $\epsilon=0.3$ (central) and $\epsilon=0.4$ (lower panel).
The remaining model parameters have been set as follows $L=100$, $\rho=0.9$}
\label{single}
\end{figure}

To measure local segregation we define the {\em node magnetization}: 
\begin{equation}
\label{magnetization}
\langle\mu\rangle=\frac{1}{N}\sum_i\frac{|n_i^B-n_i^A|}{n_i^B+n_i^A}\, ,
\end{equation}
being $n_i^A$ and $n_i^B$ the number of $A$ and $B$ agents in node $i$. Let us observe that this {index} reaches values 
close to $1$ once each node is populated by agents of the same kind and zero in the case of perfect mixing. In 
Fig.~\ref{single} we report the local magnetisation averaged over all the nodes, as a function of time for $100$ generic simulations performed using a scale free (BA) network and we can clearly observe that the system always asymptotically goes to a segregated state.

\begin{figure}[h]
\begin{center}
\includegraphics[width=11cm]{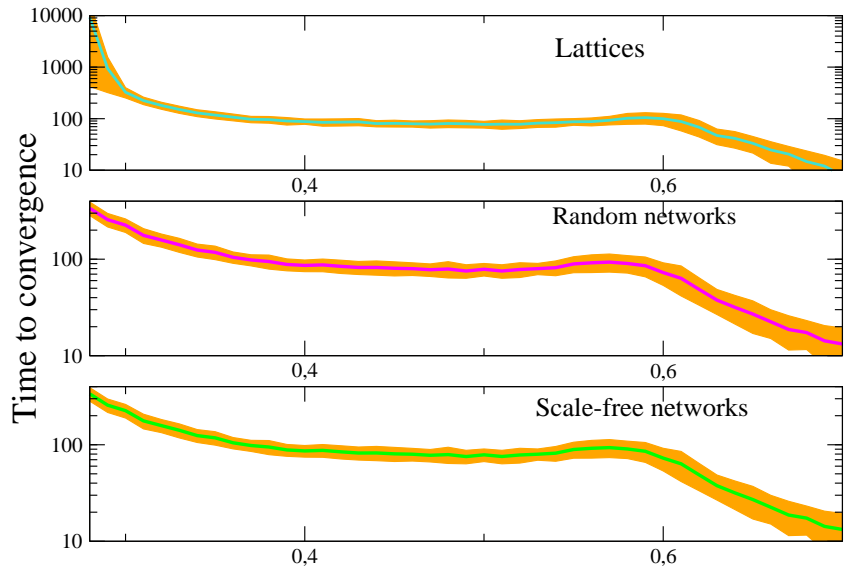}
\end{center}
\caption{Convergence time as a function of the tolerance $\epsilon$ for several topologies, upper panel $2D$ lattice, 
middle panel Erd\H{o}s-R\'enyi and lower panel (BA) scale-free networks. All the networks are composed by the same number 
$N=400$ of nodes and the parameters have been fixed to have the same average degree $\langle k\rangle=4$, 
that is a $4$--neighbourhood lattice, $p=0.01$ for the ER network and $m=2$ for the (BA) model. The solid lines represent
the average over several replicas while the shaded zones the standard variation. The remaining model parameters have 
been set as follows $L=100$, $\rho=0.9$.}
\label{time}
\end{figure}

Let us define the {\em convergence time}, to be the time needed for the system to reach the equilibrium, namely once all 
the agents are satisfied and will not move anymore. The convergence time is clearly influenced by the degree of 
connectivity for a fixed value of the tolerance; for instance in a fully connected network agents could never reach an equilibrium
if $\epsilon <1/2$, in fact for any agent in any node the fitness Eq.~\eqref{eq:happ} will take into account the whole
population and thus, because of the used initial conditions, it will result $f_i^X=1/2>\epsilon$, for $X=A,B$, hence all
agents will always be unhappy and thus they will continuously relocate themselves without achieving a stable final state.
Results reported in Fig. \ref{time} show that the convergence time is qualitatively independent from the used topology, in 
fact we can observe in all the studied networks a fast increase (exponential) of the convergence time as the tolerance 
decreases from $1$ up to $0.6$, followed by a small decreasing around $\epsilon=0.5$ and eventually start to increase once
again for $\epsilon \lesssim 0.4$. 

\begin{figure}[h]
\begin{center}
\includegraphics[width=10cm]{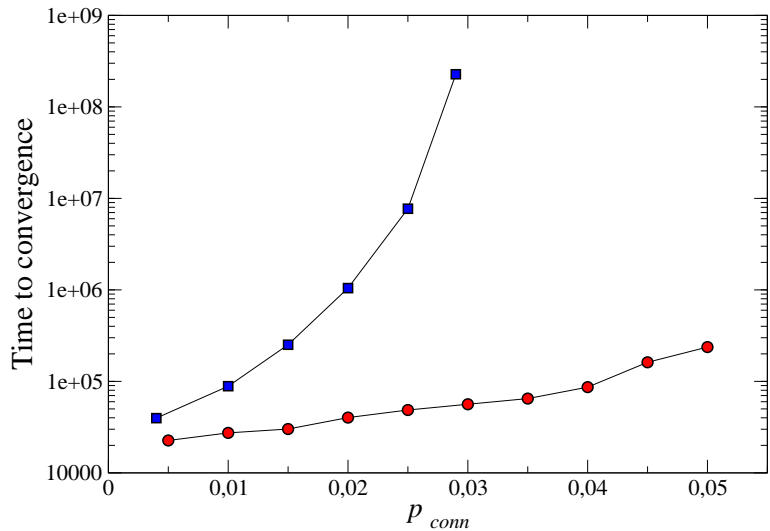}
\end{center}
\caption{Convergence time as a function of the network connectivity $p_{conn}$ for an Erd\H{o}s-R\'enyi graph with 
$N=400$ nodes, for two different values of the tolerance parameter $\varepsilon=0.3$ (red circles) and $0.4$ 
(blue squares). The model parameters have been set as follows $L=100$ and $\rho=0.9$.}
\label{pConn}
\end{figure}

To go further in this direction, we considered a random graph (constructed according to the ER model) and we tuned the 
connectivity parameter $p_{conn}$, in order to gradually increase the average degree $<k>=p_{conn}N$. As we can see in 
Figure \ref{pConn} the convergence time increases exponentially with the network average degree for $\varepsilon=0.4$. In 
the case where the system experiences a long permanence in the quasi-stationary state ($\varepsilon=0.3$) the convergence 
time grows faster than exponentially against $p_{conn}$. 

\begin{figure}[h]
\begin{center}
\includegraphics[width=16cm]{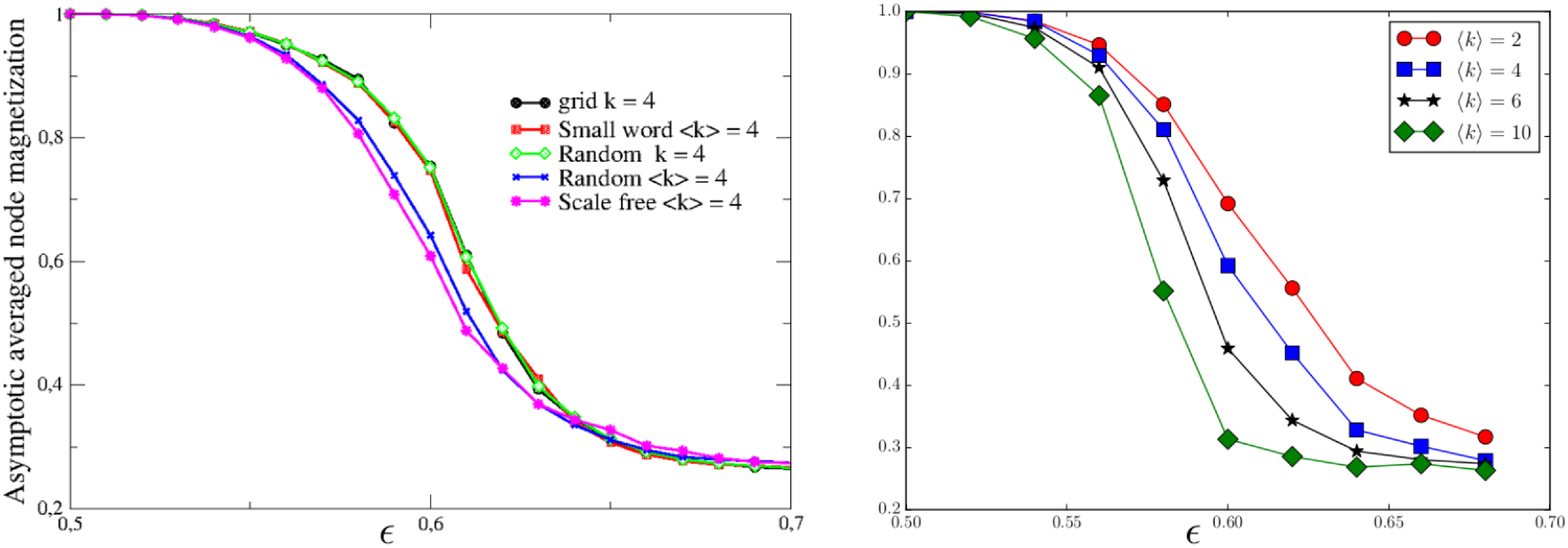}
\end{center}
\caption{Asymptotic averaged node magnetisation, Eq.~\eqref{magnetization}, as a function of the tolerance parameter. Left panel: $2D$ 
lattices (black), random networks with a fixed node degree $k=4$ (blue), WS small word networks $p=0.08$ (red), ER random 
networks with  $\langle k \rangle=4$ (yellow) and scale-free networks still with $\langle k \rangle=4$ (cyan). Each point 
is the average over $100$ realisations. Right panel: Erd\H{o}s-R\'enyi random graph with different $\langle k\rangle$. One can appreciate the qualitatively similar behaviour for all the considered cases, only the transition becomes sharper as $<k>$ increases.}
\label{mag} 
\end{figure}

We are then interested in studying the impact of the tolerance on the final outcome of the model and the possible role of 
the topology. In Fig.~\ref{mag} we report the asymptotic value of $\langle\mu\rangle$ as a function of the tolerance 
parameter $\epsilon$ for several complex network topologies where we fixed the average degree $\langle k\rangle =4$ 
(left panel) and for a ER random network varying the average degree (right panel). From this figure we see that the 
impact of the topology is very small: all curves exhibit at convergence to the same $Z$-shape. Therefore we 
can conclude that the qualitative behaviour of the system is the same for all the topologies and only the tolerance 
matters to determine a segregated state or a well mixed one.

\begin{figure}[h]
\begin{center}
\includegraphics[width=11cm]{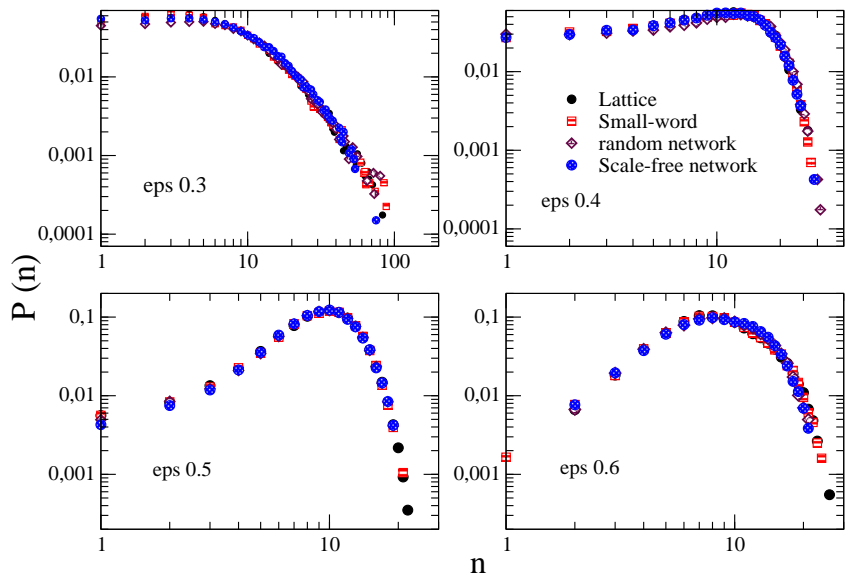}
\end{center}
\caption{Distribution of the nodes population $n=n^A+n^B$. We considered several topologies, $2D$ lattice (black circles), 
(ER) random network (brawn diamonds), (WS) small world network (red squares) and (BA) scale-free (blue circles) with the same number of nodes, $N=400$ and the same average degree $\langle k \rangle= 4$. The remaining model parameters are $L=100$, $\rho=0.9$ and several values for the 
tolerance $\epsilon=0.3$ (top left), $\epsilon=0.4$ (top right), $\epsilon=0.5$ (bottom left) and $\epsilon=0.6$ (bottom right). Each point is the average over $100$ realisations.}
\label{distri}
\end{figure}

\begin{figure}[h]
\begin{center}
\includegraphics[width=11cm]{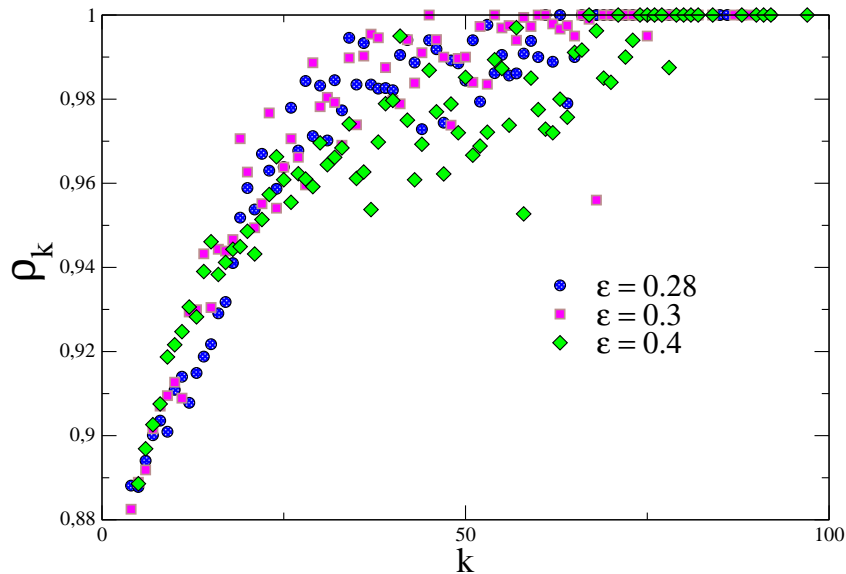}
\end{center}
\caption{Emptiness proportion, Eq.~\eqref{emptiness}, as a function of the node degree for (BA) scale-free networks with $N=400$ 
nodes and average degree $\langle k\rangle=4$. Each point is the average over $100$ realisations. One can clearly see 
that hubs are the preferred nodes for the empty places. The model parameters are $L=100$, $\rho=0.9$, $\epsilon=0.28$ 
(blue), $\epsilon=0.3$ (pink) and $\epsilon=0.4$ (green). The same output has been checked for the case when only moves 
that increase the agents happiness are allowed (solid model [12]).}
\label{degree}
\end{figure}

Les us now consider in more details the finer structure of the magnetised state arising for $\epsilon\leq 0.5$, namely 
{the population density across networks nodes.} As previously stated in the case of regular lattice, it has been found \cite{gargiulo} that for low values of the tolerance threshold the population asymptotically converges to a very heterogeneous distribution of agents in each node, despite the uniform initial distribution. In the present work we found a similar result (see Fig. \ref{distri} where we show the final distribution of 
the nodes population for different topologies), hence the heterogenous distribution is not a feature of the regular lattice
but it is a emergent outcome of the model dynamics. However in the case of complex topologies we can observe that highly 
connected nodes tend to be very often empty. This result is reported for the case of a scale free (BA) network in Fig. \ref{degree} 
where we show the average emptiness as function of the node degree:

\begin{equation}
\label{emptiness}
\langle \rho_k \rangle= \frac{1}{N_k} \sum_{i: k_i=k} \frac{L-n^A_i-n^B_i}{L}    \, ,
\end{equation}
where the sum is restricted to nodes whose degree is $k$ and $N_k$ is the total number of nodes with such degree. We 
decided to report such result only for the scale free (BA) network because this is the topology where the degree inhomogeneity is the stronger. 
It is clear that $\langle \rho_k \rangle$ increases monotonically with the degree, in some cases it reaches the value 
$1$ for nodes with very large degree (hubs). We have checked that this important result is robust with respect to the 
way agents relocate themselves, in particular we also tested the case where only moves that increase the agents happiness
are allowed (solid model \cite{DallAstaEtAl2008}) and we have found the same results, namely, emptiness increases monotonically with the degree.  

This finding can be explained using the very simple topology of a star network, where the central node - the unique hub - is linked to all the other ones that on the contrary are only linked to the central one. In this case the fitness Eq.~\eqref{eq:happ} of any agent in the central node takes into account the whole population and thus (once again because of the used initialisation) $f_{hub}^X=1/2>\epsilon$, hence no agent, whatever her type $X=A,B$, will be happy in the hub and thus they will all move away from the central node leaving it empty.

\begin{figure}[h]
\begin{center}
\includegraphics[width=11cm]{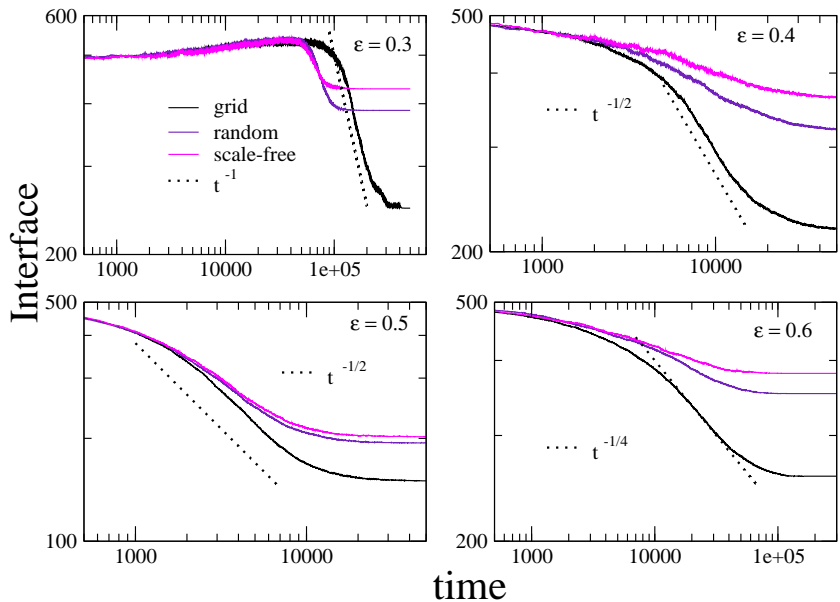}
\end{center}
\caption{Interface growth for several topologies: lattices, random and scale-free networks with same number of nodes, $N=400$, and same average degree, $\langle k \rangle = 4$, for $\epsilon=0.3, 0.4, 0.5$ and $0.6$. Each line represents the average over $100$ independent simulations. Different kinds of coarsening processes 
were found depending on the tolerance parameter. The underlying topology alters the early stage of the coarsening process, once large values of the tolerance are used, while for lower ones the phenomenon results to be topology independent.}
\label{inter}
\end{figure}

Another interesting consequence of the complex topology is related to the time evolution of the borders between the  segregation clusters, also named interface and defined as the set of nodes $i$ and $j$ such that $n^A_i \neq  n^A_j$ , $n^B_i \neq  n^B_j$ and $(n^A_i-n^B_i)(n^A_j-n^B_j)\leq 0$. Generically speaking, the driving force behind coarsening is the surface tension aiming to minimise the interface between domains as a consequence of any dynamics that promotes ordering; a similar mechanism could be acting in social systems~\cite{rmp_castellano}, and in this framework the 
Shelling model is not an exception \cite{DallAstaEtAl2008,VinkovicKirman2006}. In \cite{DallAstaEtAl2008} authors have reported results concerning a coarsening process with 
the typical size of clusters growth following the law  $t^{-1/2}$, namely an universal power law exponent independent of $\epsilon$. One expects however that this value could not describe
the coarsening process in the metapopulation framework. In \cite{gargiulo}, the authors found that for the metapopulation Schelling model on 2D regular lattice, this critical exponent achieves different values. In order the global separation process takes places, some local process inside each node should happen. This coarsening inside the coarsening could lead to smaller exponents. We found that the dynamics upon
regular lattice provides a similar exponent $-0.45\pm 0.04$ for the $\epsilon=0.4$ and $-0.48\pm 0.03$ for $\epsilon=0.5$ (see Fig.\ref{inter}), while for $\epsilon=0.6$ something interesting happens; as a consequence of the larger value of the tolerance the system falls into a metastable state characterised by the presence of many nodes each one with its own majority state where agents of both kinds do coexist. After this transient, the agents present in the minority group inside these clusters decide to emigrate, driving the system to the final state characterised
by clusters of only Red or Blue agents, separated by the remaining unidimensional clusters of mixed populations. This process takes more time (see Fig. \ref{time}) for $\epsilon = 0.6$, leading to a scaling for the interface growth of the form $t^{z}$ with $z=-0.21\pm 0.02$ (see Fig.\ref{inter}).


On the other hand, a completely different mechanism occurs for lower values of
$\epsilon\sim 0.3$. In this case the critical exponent results to be equal to
$-0.98\pm 0.21$, this is because of the long time during which the system remains in what
we have called a quasistationary state. This exponent has been previously
found in late stages of growth in concentrated non-dissipative systems \cite{siggia1979}. 
During the quasistationary state the clusters have already been formed and the
dynamics is devoted to the local arrangement of the agents to build the
towers.

The natural question now is if the above coarsening processes are robust to changes of topology. Complex
topologies change the coarsening time once only one agent is allowed per site \cite{dynamical_processes}. Our results suggest that, in the metapopulation scenario, only the late-stage phase is robust over the change of topology as we show in Fig. \ref{inter}, that is using 2D lattice, random (ER) and scale-free networks. 
For larger values of $\epsilon$, associated to faster evolution to the frozen state, the topology has an impact to the typical scale of the cluster coarsening growth, which thus results different in complex networks with respect to 2D lattice.

\section{Conclusions} \label{sec:conclusions} 

A {metapopulation} version of the Schelling segregation model has been studied over complex topologies. The main characteristics of the model have been studied in lattices, Small-Word, random and Scale-Free networks. {We have demonstrated that patterns formation in the metapopulation Schelling model are topology independent}. Therefore the segregation processes are intrinsic to the individual tolerance. The role of the topology is related to the place where patterns emerge. We have shown that hubs are the preferred places for emptiness, when empty space do exist as border between clusters, that is for low value of tolerance parameter. We demonstrated that the time to reach equilibrium is qualitatively independent 
from the used topology. The time evolution of the interface growth has also
been studied. We found different kinds of kinetics during the coarsening
process, depending on the tolerance values. The generic coarsening process in
dissipative systems have been found for $0.4 < \epsilon < 0.5$. A different
kind of process occurs for $\epsilon=0.6$ in this case the system gets trapped
for a short transient in a metastable configuration, where clusters with a
mixture of agents of both kinds transform to one dimentional
structures. Finally for lower value of the tolerance 
the coarsening process behaves as in late coarsening processes. We relate this
late stage coarsening process with local process of towers formation that
takes place during the quasistationary state. We have shown that changes in
the topologies destroy the early coarsening process occurring for large values
of tolerance, but not the late coarsening process. This special coarsening is
robust against changes of topologies. 

\section{Acknowledgements}
This paper presents research results of the Belgian Network DYSCO (Dynamical Systems, Control, and Optimisation), funded by the
Interuniversity Attraction Poles Programme, initiated by the Belgian State, Science Policy Office.


\begin{thebibliography}{*}
\bibitem{CutlerGlaeser1997} Cutler, D. and Glaeser, E.  Are Ghettos Good or Bad?. {\it Quart. J. Econ.} {\bf112}: 827-872. (1997)
\bibitem{FagioloEtAl2007} Fagiolo, G., Valente, M. and Vriend, N.  Segregation in networks. {\it J. Econ. Behav $\&$ org} {\bf64}:316-336. (2007) 
\bibitem{FreyFarley1996} Frey, W. and Farley, R.  Latino, Asian and black segregation in U.S. Metropolitan areas: Are multiethnic metros different? {\it Demography} {\bf 33}:35-50. (1996)
\bibitem{Schelling1969} Schelling, T.C.  Models of Segregation. {\it American Econ. Rev.} {\bf 59}: 488-493. (1969)
\bibitem{Schelling1971} Schelling, T.C.  Dynamic models of segregation. {\it J. Math. Soc.} {\bf 1}: 143-186. (1971)
\bibitem{PancsVriend2007} Pancs, R. and Vriend, N.  Schelling's spatial proximity model of segregation revisited. {\it J. of Public Economic} {\bf 91}:1-24. (2007)
\bibitem{Gauvin2013} Gauvin L, Vignes A and Nadal J-P. Modeling urban housing market dynamics: Can the
socio-spatial segregation preserve some social diversity? J. Econ, Dynamics $\&$ Control {\bf 37}: 1300 (2013)
\bibitem{Hanski1999} Hanski, I., Metapopulation ecology. Oxford University Press, (1999)
\bibitem{DurettYuan2014} Durrett, R. and Zhang, Y.  Exact solution for a metapopulation version of Schelling’s model. {\it Proc Natl Acad Sci USA }{\bf 111(39)}:14036-14041.  (2014)
\bibitem{RogersMcKane2012} Rogers, T. and McKane, A.  Jamming and pattern formation in models of segregation. {\it Phys. Rev. E} {\bf 85}:041136 (1-5). (2012)
\bibitem{VinkovicKirman2006} Vinkovi\' c, D. and Kirman, A.  A physical analogue of the Schelling model. {\it Proc Natl Acad Sci USA} {\bf 103(51)}:19261-19265. (2006)
\bibitem{DallAstaEtAl2008} Dall'Asta, L., Castellano, C. and Marsili, M.  Statistical physics of the Schelling model of segregation. {\it J. Stat. Mech} {\bf 7}(L07002):1-10. (2008)
\bibitem{gargiulo} Gargiulo F., Gandica Y. and Carletti T. Urban skylines from Schelling model. ArXiv:1505.00429 (physics.soc ph).
\bibitem{UniversalityBC} Fortunato, S. Universality of the Threshold for Complete Consensus for the Opinion Dynamics of Deffuant et al. International Journal of Modern Physics C {\bf 15.09}: 1301-1307 (2004).
\bibitem{OpDynNetworks} Gargiulo, F. and S. Huet. Opinion dynamics in a group-based society. EPL {\bf 91.5}: 58004 (2010).
\bibitem{OpDynHolme} Holme, Petter, and Mark E. J Newman. Nonequilibrium phase transition in the coevolution of networks and opinions. Physical Review E {\bf 74.5}: 056108 (2006).
\bibitem{Watts_Stogatz} D. J. Watts and S. H. Strogatz. Collective dynamics of 'small-world' networks. Nature {\bf 393} 409 (1998).
\bibitem{ER} Erd\H{o}s, P., and R\'enyi, A.. On the evolution of random graphs. Publ. Math. Inst. Hung. Acad. Sci {\bf 5} :17-61 (1960).
\bibitem{maslov} S. Maslov, K. Sneppen, U. Alon, in: S. Bornholdt, H.G. Schuster (Eds.), Handbook of Graphs and Networks, Wiley-VCH and Co. Weinheim (2003).
\bibitem{BA}Barabasi L. and Albert R. Statistical mechanics of complex networks. Rev. Modern Physics, {\bf 7}:47-97 (2002).
\bibitem{rmp_castellano} C. Castellano, S. Fortunato and V. Loreto, Statistical physics of social dynamics. Rev. Mod. Phys. {\bf81}: 591 (2009).
\bibitem{siggia1979} Siggia E., Late stages of spinodal descomposition in binary mixtures. Phys. Rev. A, {\bf 20} (1979).
\bibitem{dynamical_processes} A. Barrat, M. Barth\'{e}lemy and A. Vespignani. Dynamical Processes on Complex Networks, Cambridge University press (2008). 
\end{thebibliography}
\end{document}